\documentclass[12pt,preprint]{aastex}
\usepackage{epsf}

\begin{document} 

\title{Dusty Rings: Signposts of Recent Planet Formation}

\author{Scott J. Kenyon}
\affil{Smithsonian Astrophysical Observatory, 60 Garden Street, 
Cambridge, MA 02138, USA; e-mail: skenyon@cfa.harvard.edu}
\author{and}
\author{Benjamin C. Bromley}
\affil{Department of Physics, University of Utah, 201 JFB, Salt Lake City, 
UT 84112, USA; e-mail: bromley@physics.utah.edu}

\begin{abstract}

Many nearby stars are surrounded by a bright ring or disk of cold dust. 
Our calculations show that these disks and rings of dust are signposts 
of recent planet formation. Bright rings appear because dust associated 
with the formation of a planet absorbs and scatters light from the 
central star. The calculations explain the rings observed so far 
and predict that all nascent solar systems have dusty rings. 

\end{abstract}

\subjectheadings{planetary systems -- solar system: formation -- 
stars: formation -- circumstellar matter}

\section{INTRODUCTION}

Every planetary system forms from a thin disk of gas and dust in orbit 
around a young star.  In the planetesimal theory, planets grow from 
collisions and mergers of smaller bodies, planetesimals, embedded
in the disk.  Protoplanets with radii of 100 km or more stir up the 
remaining planetesimals along their orbits.  A cascade of collisions 
among rapidly moving planetesimals produces a ring of dust grains, 
which slowly disappears as protoplanets grow into planets.  This 
entire process can lead to a solar system similar to our own 
\citep{lis93,man00}.

Recent observations support this general picture.  The dusty disks 
around many nearby stars are as large or larger than our solar system 
\citep{koe01,wei02}.  The dusty ring around the nearby 10 Myr old star, 
HR 4796A, 
has a thickness of $\sim$ 15 AU and lies $\sim$ 70 AU from its central 
star \citep{jay98,koe98,sch99,gre00,tel00}.  The rings or partial rings around 
$\epsilon$ Eridani, Vega, and other older stars have similar dimensions 
\citep{den00,koe02,wil02}.  

There have been few detailed numerical calculations of planet formation 
for comparison with these modern observations. Some calculations 
explore the early stages of planetesimal growth in a small range of 
disk radii \citep{gre84,ws93}.  These models follow the development 
of a single large planet, typically the Earth or Jupiter.  Others use 
$n$-body simulations to investigate the last stages of planet formation, 
when large bodies coalesce to form a few planets \citep{lis96,lev98,cha01}. 

To simulate the formation of an entire planetary system, the calculation 
must span a large range of disk radii. Extending the calculations over 
a large fraction of the disk allows simulation of the diversity of 
observable phenomena in extrasolar planetary systems. Supercomputers 
now allow such {\it multiannulus} planetesimal calculations covering a 
decade or more in disk radius \citep{spa91,wei97,kor01}.  
The multiannulus calculations we discuss predict the behavior of 
a planet-forming disk during the early and intermediate stages
of planet formation. The models yield images for comparison with
observations of disks around nearby stars.

\section{THE MODEL}

To calculate planet growth, we adopt the \cite{saf69} formalism,
which treats planetesimals as a statistical ensemble of bodies with 
a distribution of masses and velocities orbiting a central star.
Because direct orbit integrations of 10$^{20}$ or more planetesimals 
are not possible, this statistical approach is essential.  As long 
as planetesimals are numerous and gravitational interactions are small, 
statistical calculations reproduce the results of direct orbit 
calculations \citep{kok00}.

To evolve the mass and velocity distributions in time, we solve the 
coagulation and Fokker-Planck equations for bodies undergoing inelastic 
collisions, drag forces, and long-range gravitational forces 
\citep{kb02,wei97}.
We approximate collision rates as $n \sigma v f_g$, 
where $n$ is the number density, $\sigma$ is the geometric cross-section, 
$v$ is the relative velocity, and $f_g$ is the gravitational focusing factor.  
For small planetesimals with radii $\le$ 1--10 km, $f_g$ = 1.
For larger planetesimals, the gravitational cross-section exceeds
the geometric cross-section and $f_g > $ 1. We use the ratio of the
center-of-mass collision energy $E_c$ to the sum of the tensile strength
$S_0$ and the gravitational binding energy of the merged pair $E_g$ to 
assign collision outcomes. Collisions with $E_c/(E_g +S_0)\ll $ 1 produce 
mergers with little dust; $E_c/(E_g + S_0) \sim$ 1 yields mergers with 
dust; $E_c/(E_g + S_0) \gg$ 1 yields only dust.  This algorithm matches 
data for simulated laboratory collisions and observations of the asteroid 
belt in our solar system \citep{dav85,dav94}.  We then compute velocity 
changes from gas drag, dynamical friction and viscous stirring 
\citep{kb02,oht02}.  Dynamical friction transfers kinetic energy from 
large bodies to small bodies and drives a system to energy equipartition.  
Viscous stirring transfers angular momentum between bodies and increases 
the velocities of all planetesimals.  

Our numerical calculations begin with 1--1000 m planetesimals in 64 
concentric annuli at distances of 30--150 AU from a 3 $M_{\odot}$ star.
The total mass in planetesimals is $M_0$. We divide the initial 
continuous mass distribution of planetesimals into a differential
mass distribution of 30 mass batches with spacing $m_{i+1}/m_i$  = 2 
between successive mass batches. We add mass batches as planetesimals 
grow in mass.  Each planetesimal batch begins with a nearly circular 
orbit with eccentricity $e_0$ and inclination $i_0 = e_0/2$.  

\section{RESULTS}

Figures 1--3 summarize results for a model with $M_0$ = 100 $M_\oplus$
(1 $M_\oplus$ = $6 \times 10^{27}$ g is the mass of the Earth) and $e_0$ 
= $10^{-5}$.  This initial mass is appropriate for the `minimum mass
solar nebula,' the minimum amount of solid material needed for the
planets in our solar system \citep{hay81,wei77}. The adopted $e_0$ 
is a reasonable 
equilibrium value for 1--1000 m objects.  This model assumes icy 
bodies with a mass density, $\rho_p$ = 1.5 g cm$^{-3}$, and a tensile 
strength comparable to terrestrial snow, $S_0 = 10^6$ erg g$^{-1}$.
These parameters are similar to those adopted for calculations of
planet formation in the Kuiper Belt of our solar system \citep{ken02}.

We separate the growth of planetesimals into three stages. When 
planetesimals are small, they have small geometric cross-sections
and slow growth rates.  This slow growth erases the initial 
conditions, including $e_0$ and the initial mass distribution. 
Slow growth ends when the gravitational cross-sections of the
largest objects exceed their geometric cross-sections. This
`gravitational focusing' enhances collision rates by a factor of 
100--1000.  A few of the largest objects begin `runaway growth';
their radii grow rapidly from $\sim$ 10 km to $\sim$ 300 km. During 
the runaway, dynamical friction and viscous stirring increase the 
eccentricities of the 1 m to 1 km radius bodies from $e \sim 10^{-4}$ 
to $e \ge 10^{-3}$.  Collisions between these objects then begin 
to produce substantial amounts of dust.  
Gravitational focusing factors diminish and runaway growth ends.  
As the largest bodies grow slowly to 1000--3000 km sizes, they 
continue to stir up the smaller bodies and produce more and more dust.  
Eventually, nearly all of the small bodies are gone and dust production 
slows down.  Radiation pressure from the central star removes 1 $\mu$m 
dust grains; Poynting-Robertson drag pulls 1--100 $\mu$m dust grains 
into the central star.  These processes remove dust on short timescales,
$\sim$ 10 Myr or less \citep{art89,bac95}. 

Figure 1 shows the evolution of the eccentricities $e$ of planetesimals
at the inner edge of the disk (30--36 AU).  All bodies begin with
the same eccentricity.  As objects grow from $\sim$ 1 km to $\sim$
1000 km, dynamical friction maintains a roughly power law eccentricity
distribution for the largest bodies.  Viscous stirring steadily 
increases the eccentricities of the smallest bodies.  When $e \ge$
0.01, collisions between small bodies begin to produce dust.
Continued stirring yields larger $e$ and more dust, until the small 
bodies are gone.  The amount of dust then steadily decreases with time.

Figure 2 shows snapshots of the time evolution of the largest object 
in each annulus.  For most of the evolution, the growth timescale for 
planetesimals is roughly $P/\Sigma \propto A^{-3}$ where $P$ is the 
orbital period for heliocentric distance $A$ and $\Sigma$ is the surface 
density of planetesimals \citep{lis87}.  Planetesimals thus grow first 
at the inner 
edge of the disk.  It takes less than 1 Myr to produce 10 km bodies
in the inner disk (30 AU) and $\sim$ 100 Myr to produce 10 km bodies in the 
outer disk (150 AU).  Runaway growth leads to rapid production of 100--300 km 
bodies on timescales of 5 Myr in the inner disk and 100--200 Myr in 
the outer disk.  A second slow growth phase produces objects ranging
in size from Pluto (1000 km radius) to Jupiter's moon Ganymede ($\sim$
2500--3000 km radius).  It takes 20--30 Myr to produce these objects 
in the inner disk and close to 2 Gyr to make large objects in the outer 
disk.

To visualize dust formation in a planet-forming disk, we use the geometric 
optics limit to derive the optical depth $\tau$ of particles in the grid
\citep{ken99}.  
For each object with radius $r_j$ and space density $n_j$ in a single annulus 
of width $\delta A_i$, the optical depth is $\tau_{ij} = 2 \pi n_j r_j^2 
\delta A_i$.  We derive $\tau_i$ for each annulus as the sum of $\tau_{ij}$ 
over all objects.  The relative brightness of each annulus is then 
proportional to the solid angle of each annulus on the sky as seen 
from the central star, $L_i/L_{\star}$ = $\tau_i H_i / A_i$, where 
$H_i$ is the scale height of the dust in annulus $i$.  This result 
assumes $\tau_i \lesssim$ 0.1 at all wavelengths, which we verify 
for each timestep.  The brightness per unit surface area of the disk 
follows from $L_i$ and the surface area of each annulus, $\mu_i \propto L_i 
/ 2 \pi A_i \delta A_i$.  The total luminosity is the sum of $L_i$ for 
all annuli, $L_d/L_{\star} = $ $\sum \tau_i H_i / A_i$, where
$L_d/L_{\star}$ is the brightness of the disk relative to the brightness
of the central star.

Figure 3 shows nine snapshots of the disk viewed along the rotational
axis.  In each image, the bright point in the center is the star.
The sequence begins with a faint disk in the upper left corner 
at $t$ = 0.  From $t$ = 0 to $t$ = 1 Myr, the inner disk fades slightly
as planetesimals grow from 1 km to $\sim$ 30--50 km.  By $t$ = 3 Myr, the 
largest objects have radii of 200--300 km.  Collisions between the
smallest objects produce modest amounts of dust; the inner disk
brightens.  In the middle panels, the inner disk brightens dramatically
as collisions between the smallest objects produce more and more dust.
Outer regions of the disk fade and then begin to brighten as large objects
begin to form and stir up smaller objects.  The contrast in brightness
between the bright ring (shown in blue) and the rest of the disk is a
factor of 30--100, making it readily observable.  In the lower panels, 
the bright ring moves outwards.  Throughout the expansion, this ring
often contains narrow dark gaps that are local minima in the dust
production rate.  Dust disappears in the inner disk;
collisions produce more dust in the outer disk.  By $t$ = 2.5 Gyr, 
the bright ring reaches the outer boundary of the disk and the entire 
disk starts to fade and becomes unobservable \citep{hab01,spa01}.

In the animation of Figure 3 included in the electronic version of the
paper, planet formation appears as a set of three waves propagating 
outward through the disk.
\footnote{This animation and an animation for
Figure 4 are also available at 
http://cfa-www.harvard.edu/~kenyon/pf/sp/movies.html}.
Slow growth from 1 km to $\sim$ 100 km produces a
dark wave that lowers the brightness of the disk.  Dust formed during
runaway growth produces a bright wave that appears as a series of bright, 
narrow, concentric rings in the disk.  
The disk is brightest at $t \sim$ 100 Myr, when $L_d/L_{\star} \sim$ 
$10^{-3}$. Finally, the disappearance of the dust grains yields a second 
dark wave that signals the last phases of planet formation in the disk. 
During this last phase, large bodies remaining in the disk will coalesce
and form planets.  

To test the robustness of these results, we calculated a series of models 
with a variety of initial conditions.  We changed the initial eccentricity 
$e_0$, mass density $\rho_p$ and tensile strength $S_0$ of the bodies, and 
the total mass $M_0$ in the planetesimals.  Most of these parameters change 
the timescale but do not change the character of the evolution 
\citep{wet89,kl99b}.  Because 
gravitational focusing is less effective at larger planetesimal velocities, 
larger initial eccentricities, $e_0 > 10^{-5}$, delay runaway growth and 
slow the progress of the bright wave through the disk.  The delay is 
roughly $(e_0 / 10^{-5})^{1/2}$.  Planetesimals with smaller mass densities
have larger cross-sections and thus grow faster.  A larger initial mass in
planetesimals increases the collision rate and shortens the evolution time.
The evolution time is inversely proportional to the initial mass,
$t \propto M_0^{-1}$, and to the planetesimal mass density,
$t \propto \rho_p^{-2/3}$.  Finally, changes in the tensile strength $S_0$
affect the growth time and the final mass of planetesimals.  Weaker bodies
with $S_0 < 10^6$ erg g$^{-1}$ produce more dust in each collision and
disrupt at smaller collision velocities.  More dust during linear growth 
enhances dynamical friction and shortens the timescale to runaway growth.
More dust after runaway growth speeds the exhaustion of smaller bodies
and limits the growth of the largest bodies.  For $S_0$ = $10^2$ to $10^6$
erg g$^{-1}$, the net change in evolution time is $\sim$ 10\% to 30\%.
The variation in the size of the largest object is 
$r_{max} \propto S_0^{\rm -0.20 ~ to ~ -0.25}$.

The evolution is more sensitive to stochastic processes.  During runaway
growth,  random fluctuations in the collision rate can produce a single
large body which grows much more rapidly than anything around it. 
By robbing other bodies of material, this runaway body slows the 
growth of large objects nearby.  Stirring by the runaway object
leads to more dust on shorter timescales. These runaways always produce
bright dust rings.

To illustrate this process, Figure 4 shows disk snapshots for 
a model with $M_0$ = 100 $M_\oplus$, $e_0$ = $10^{-5}$, 
and $S_0$ = $10^4$ erg g$^{-1}$.  The electronic version of the paper
contains an animation of the surface brightness evolution. For $t <$ 
1 Myr, this model follows the evolution of the model in Figure 3.
At $t \approx$ 1 Myr, single runaway bodies in two adjacent annuli
begin to grow much more rapidly than anything else. They grow to
1000 km in $\sim$ 3 Myr.  Stirring by these runaway bodies produces 
copious amounts of dust, which we see as two bright rings at 36 AU 
and at 40 AU.  For $t$ = 10--100 Myr, collisions in the inner disk 
exhaust the supply of small bodies. At $t \approx$ 100 Myr, a 
fluctuation produces another runaway body at 100 AU. Its rapid growth 
produces another bright dust ring outside the growing inner ring.  
As collisions deplete small bodies in the inner disk and begin to 
fragment small bodies in the outer disk, both of these rings and 
several fainter and narrower rings propagate outward through the disk.

\section{SUMMARY}

In our calculations, multiple rings tend to form when planets grow 
rapidly. Rapid growth occurs when the mass density of colliding bodies
$\rho_p$ is small and when the total mass in planetesimals $M_0$ is 
large. Bodies with low tensile strength also promote rapid growth and
multiple ring production.  Calculations with $S_0 \le 10^4$ 
erg g$^{-1}$ produce multiple rings more often than do calculations 
with $S_0 \sim 10^6$ erg g$^{-1}$.  Multiple ring production is 
insensitive to $e_0$, the initial mass distribution, and other 
initial conditions.

Our calculations demonstrate that planet formation produces copious 
amounts of dust.  The dust production rate ranges from 
$\sim 10^{18}$ g yr$^{-1}$ to $10^{21}$ g yr$^{-1}$.
This dust absorbs and reradiates stellar energy with a relative
luminosity of $L_{\rm dust}/L_{\star}$ $\sim 10^{-5}$ to $10^{-3}$,
comparable to observed luminosities for dusty disks surrounding 
nearby stars \citep{hab01,spa01}. Our dust formation timescales
of 10--100 Myr are comparable to the ages of nearby stars with dusty
disks \citep{hab01,son00,spa01}.  Dust first forms in large quantities when 
the largest bodies reach sizes of $\sim$ 1000 km.  Dust disappears 
when disruptive collisions exhaust the supply of $\sim$ 1 km bodies
and radiative processes remove dust from the ring.  Thus, dust is 
concentrated in concentric rings which propagate outward through the 
disk as a function of time.  The outer edge of each ring marks the 
location where 1000 km objects are just starting to form; the inner 
edge marks the location where collisions have exhausted the supply 
of $\sim$ 1 km bodies and dust has disappeared.  Thus, dusty rings are 
signposts for recent formation of 1000 km or larger planets surrounding 
a star.

\vskip 6ex

We acknowledge a generous allotment, $\sim$ 500 cpu days, of
computer time on the Silicon Graphics Origin-2000 `Alhena' at the
Jet Propulsion Laboratory through funding from the NASA Offices of
Mission to Planet Earth, Aeronautics, and Space Science.  Advice
and comments from J. Brauman and M. Geller greatly improved our
presentation.  R. Mackey of the JPL supercomputing group assisted
with the animations of Figure 3 and Figure 4.

\clearpage


\begin{figure}
$   $
\hskip 12ex
\plotone{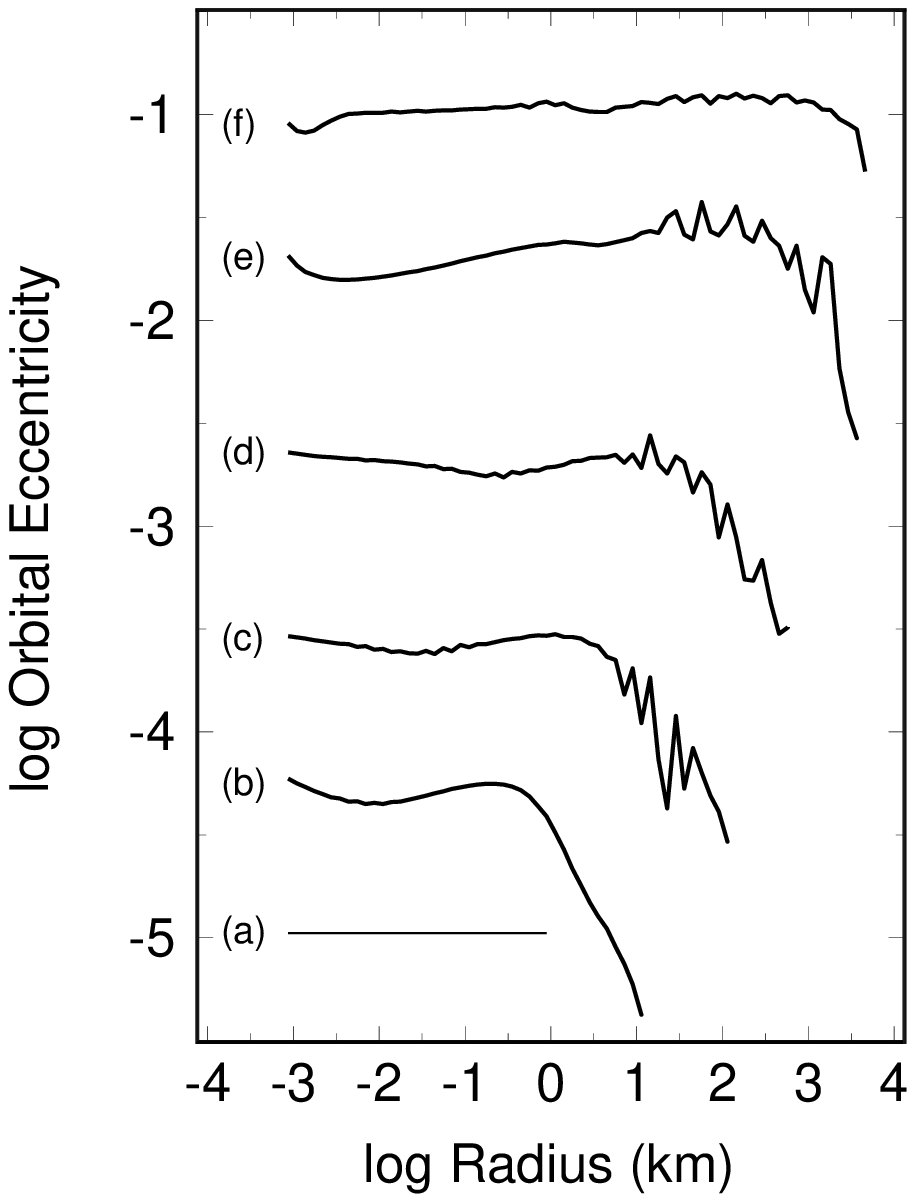}
\caption
{Eccentricity distribution for planetesimals at 30--36 AU in a
planet-forming disk. Each curve shows the distribution at different times
in the evolution, (a) $t$ = 0, (b) $t$ = 1 Myr, (c) $t$ = 3 Myr,
(d) $t$ = 10 Myr, (e) $t$ = 100 Myr, and (f) $t$ = 2 Gyr.}
\end{figure}
\clearpage

\begin{figure}
\plotone{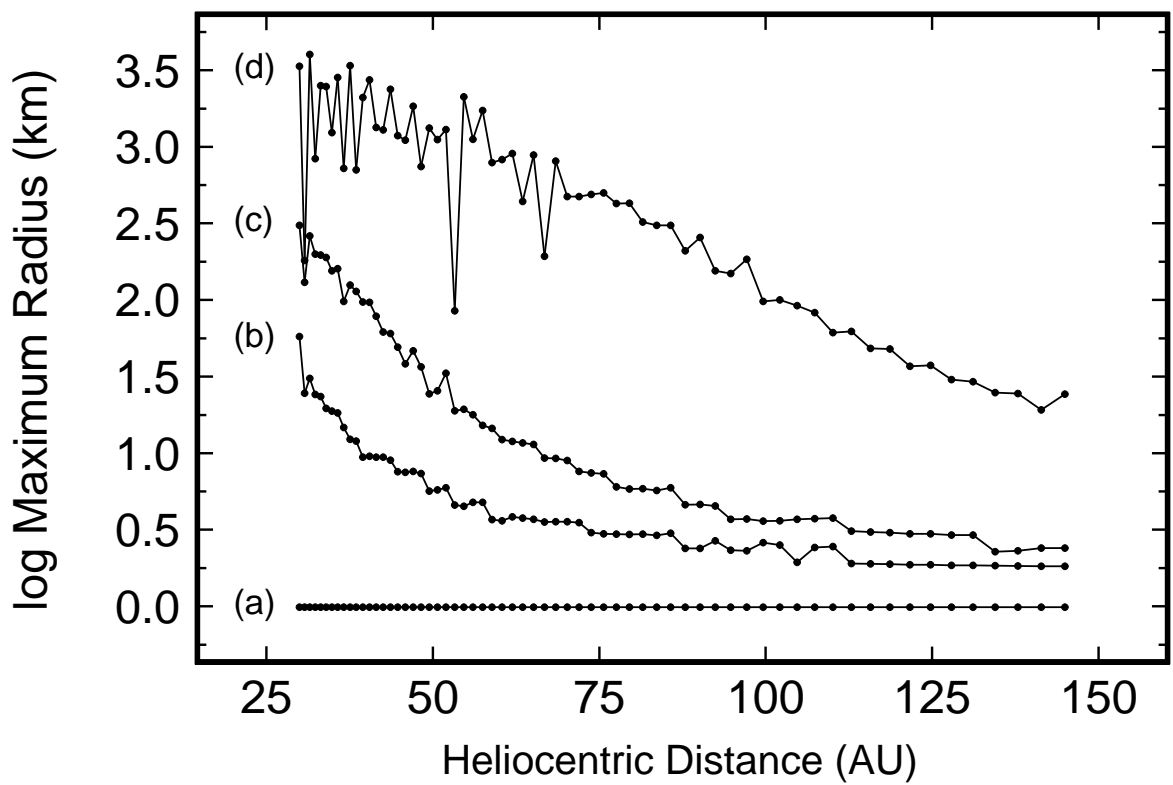}
\caption
{Mass of the largest object at (a) $t$ = 0, (b) $t$ = 1 Myr,
(c) $t$ = 5 Myr, and (d) $t$ = 125 Myr.  Planets grow rapidly at the
inner edge of the disk and more slowly at the outer edge. The growth
rate is proportional to $A^3$, where $A$ is the heliocentric distance.
Rapid growth in several radial zones can rob material from bodies in
neighboring zones, producing pronounced dips in the distribution for
$t >$ 10--20 Myr.}
\end{figure}
\clearpage
 
\begin{figure}
\caption
{Nine false-color snapshots of a planet-forming disk.  Bright blue is 
brightest; dark red is dimmest.  The top images show how the inner disk 
fades as 1--1000 m planetesimals grow into 10--100 km objects from
$t = 0$ (upper left panel) to $t$ = 3 Myr (upper middle panel) and
$t$ = 14 myr (upper right panel).  The middle panels show images
at $t$ = 30 Myr, $t$ = 100 Myr, and $t$ = 300 Myr.  The outer disk
fades as 1000 km and larger planets form in the inner disk. The lower 
sequence shows images at $t$ = 600 Myr, $t$ = 1 Gyr, and $ t$ = 2 Gyr.
The inner disk fades as collisions exhaust the supply of planetesimals.
Bright rings containing distinct dark gaps form in the outer disk, 
where mergers produce 1000 km and larger objects.}
\end{figure}
\clearpage
 
\begin{figure}
\caption
{As in Fig 3 for a planet-forming disk where stochastic processes
produce two rings which propagate out through the disk.  Narrow
dark gaps in the rings indicate local minima in the dust production 
rate. The top three panels show the disk at $t$ = 0 Myr, 3 Myr, and 
10 Myr.  The middle three panels
show the disk at $t$ = 30 Myr, 100 Myr, and 300 Myr. The bottom three
panels show the disk at $t$ = 400 Myr, 1 Gyr, and 2 Gyr.}
\end{figure}
\clearpage

\end{document}